# Performance Evaluation of QoS Parameters in Dynamic Spectrum Sharing for Heterogeneous Wireless Communication Networks

Kaniezhil.R[1], Chandrasekar.C[2] and NithyaRekha.S[3]

[1] Full-Time Research Scholar, Department of Computer Science, Periyar University,
Salem, TamilNadu-636011, India
*kaniezhil@yahoo.co.in*

[2] Associate Professor, Department of Computer Science, Periyar University,
Salem, TamilNadu-636011, India
*ccsekar@gmail.com*

[3] Full-Time Research Scholar, Department of Computer Science, Periyar University,
Salem, TamilNadu-636011, India
*rekhasiva24@gmail.com*

**Abstract**
Cognitive radio nodes have been proposed as means to improve the spectrum utilization. It reuses the spectrum of a primary service provider under the condition that the primary service provider services are not harmfully interrupted. A cognitive radio can sense its operating environment's conditions and it is able to reconfigure itself and to communicate with other counterparts based on the status of the environment and also the requirements of the user to meet the optimal communication conditions and to keep quality of service (QoS) as high as possible. The efficiency of spectrum sharing can be improved by minimizing the interference. The Utility function that captures the cooperative behavior to minimize the interference and the satisfaction to improve the throughput is investigated. The dynamic spectrum sharing algorithm can maintain the quality of service (QoS) of each network while the effective spectrum utilisation is improved under a fluctuation traffic environment when the available spectrum is limited.

## 1. Introduction

It is commonly believed that there is a spectrum scarcity at frequencies that can be economically used for wireless communications. By recent studies of FCC, it shows that the scared spectrum can be well utilized and unused spectrum ie 'white spaces' can be utilized by the secondary users with the advance technology of Cognitive Radio (CR)[2] to implement the opportunistic spectrum sharing.

However, as noted by the FCC, there are large portions of allotted spectrum that are unused when considered on a time and geographical basis. There are portions of assigned spectrum that are used only in certain geographical areas and there are some portions of assigned spectrum that are used only for brief periods of time. Studies have shown that even a straightforward reuse of such "wasted" spectrum can provide an order of magnitude improvement in available capacity.

Thus the issue is not that spectrum is scarce – the issue is that most current radio systems do not utilize technology to effectively manage access to it in a manner that would satisfy the concerns of current licensed spectrum users. Cognitive radio [2], [3], [4] is currently considered as one of the most promising solutions to the aforementioned scarcity problem by enabling a highly dynamic, device-centric spectrum access in future wireless communication systems.

A CR can adapt the operation parameters of its radio (frequency band, modulation, coding etc) and its transmission or reception parameters on the fly based on cognitive interaction with the wireless environment in which it operates. CR will lead to a revolution in wireless communication with significant impacts on technology as well as regulation of spectrum usage to overcome existing barriers.

Cognitive radio not only adapts to the available spectrum but it also shows the better Qos and the channel conditions that satisfies the requirement of the effective performance of the bandwidth.

Cognitive Radio is an emerging technology provides an way to efficient way for better utilization of the unused spectrum. Spectrum allocated to the primary users is not used fully at all instances of times. Hence, the number of trying to use this unused licensed spectrum is increasing enormously. So, the idea is that the sensing the unused or empty frequencies of the primary users and that can be accommodated to some other unlicensed (Secondary) users. This makes the efficient utilization of the available spectrum. This can be achieved by using the Cognitive radio to identify and used to allocate the unused spectrum bandwidth that can allocate dynamically by changing their parameters keeping in view the QoS requested by the secondary user or simply the application, without interfering with the primary users.

The techniques developed to date for the enhancement of heterogeneous networks concentrate on improving their accessibility and QoS. Numerical simulation results demonstrate that Throughput and Spectrum Efficiency of networks employing dynamic spectrum sharing are much better than those of networks employing fixed allocation, especially for networks under heavy traffic load, when spectrum is limited.

I have already proposed the spectral efficiency in my previous work that the call arrival rate vs spectral efficiency[1]. The remaining work ie Performance of QoS based on spectrum sharing using CR nodes is carried out in the present work.

The paper is organized as follows; Section II and III defines cognitive radio and proposes a system model approach for its implementation. In section IV, Performance analysis has been investigated to improve the system efficiency. Section V presents the simulation results and implementation issues. Finally, conclusions are presented in Section VI.

## 2. Cognitive Radio

2.1 Introduction

A "Cognitive Radio" is a radio that is able to sense the spectral environment over a wide frequency band and exploit this information to opportunistically provide wireless links that best meet the user communications requirements. CR provides the real time interaction with its environment. This provides the way to dynamically adapt to the dynamic radio environment and the radio analyzes the spectrum characteristics and changes the parameters among the users that share the available spectrum. With the approach to solve the issue of scarcity of available radio spectrum, the Cognitive radio technology is getting a significant attention [4]-[6].

The primary feature of cognitive radio is the capability to optimize the relevant communication parameters given a dynamic wireless channel environment. Since cognitive radios are considered lower priority or secondary users of spectrum allocated to a primary user, a fundamental requirement is to avoid interference to potential primary users in their vicinity. On the other hand, primary user networks have no requirement to change their infrastructure for spectrum sharing with cognitive networks.

Therefore, cognitive radios should be able to independently detect primary user presence through continuous spectrum sensing. In general, cognitive radio sensitivity should outperform primary user receiver by a large margin in order to prevent what is essentially a *hidden terminal problem*. This is the key issue that makes spectrum sensing very challenging research problem.

2.2 Cognitive Radio Parameters

The Cognitive Radio system must relate the performance objectives to the transmission parameters and the environmental parameters in order to reach at an optimized solution. While defining the list of parameters we make a compromise between the large time scale, system level parameters and the small time scale, transmission level parameters.

Table I shows the transmission parameters used in this paper to generate a utility function.

TABLE I

TRANSMISSION PARAMETER LIST

| *Parameter Name* | *Symbol* | *Description* |
|---|---|---|
| Transmit Power | P | Transmission Power |
| Modulation Type | MT | Type of Modulation |

The available system parameters should be defined as decision variables for evolutionary algorithms calculating generating utility functions. Table II shows the Environmental parameters used in this paper to generate a utility function.

TABLE II

ENVIRONMENTALLY SENSED PARAMETER LIST

| Parameter Name | Symbol | Description |
|---|---|---|
| Bit Error Rate | BER | Number of bit errors divided by the total number of transferred bits during a studied time interval. |
| Signal-to-Interference Noise Ratio | SINR | Ratio of the received strength of the desired signal to the received strength of undesired signals (noise and interference). |
| Noise power | $N_O$ | Magnitude in decibels of the Noise Power |

The BER parameter value depends on several channel characteristics, including the noise level and transmit power. Environmental Parameters inform the system of the surrounding environmental characteristics. SBAC Algorithms is chosen for the allocation algorithm due to their fast convergence.

2.3 Utility Functions

The system performance indexes are described in terms of utility functions. The actual results should take balance of these utility functions, which can meet the QoS requirements and improve the performance.

Utility functions are defined individually considering the current user's QoS specifications. This implies to the existence of a trade-off among the parameters for a particular channel. This is analyzed by the corresponding weights assigned by the user to each of them. This is actually very useful in our decision-making process and provides with a variety of solutions for the best optimization of a problem.

Four performance measures of Data Transmission rate, Propagation Delay, Spectral Efficiency and Throughput are considered in this paper and the utility functions are designed as in Table III:

TABLE III

UTILITY FUNCTIONS

| Performance Metrics |
|---|
| Low Propagation Delay |
| Minimize RTT |
| Maximize Throughput |
| Maximize Spectral Efficiency |

Using the objectives in Table III as sole inputs to the utility functions will not suffice. It is ambiguous to have the system minimize power consumption while also minimizing BER. Thus, the objectives must also contain a quantifiable rank representing the importance of each. This will allow the utility function to characterize the trade-offs between each objective by ranking the objectives in order of importance. Several approaches exists for determining the preference information of a set of objectives.

## 3. System Model

We consider the spectrum sharing among multiple service providers, they belong to the licensed bands. We assume that there are a number of primary and secondary users communicating with their partners simultaneously. Here, the term "user" will be used broadly where it can be a mobile node or base station in a distributed networks. Simultaneous communications among users (i.e., both primary and secondary users) will interfere with each other. The entities we will work with are communication links each of which is a pair of users communicating with each other. We will refer to communication links belonging to secondary networks as secondary links. We will also consider the interference constraints at the receiving nodes of primary networks which will be referred to as primary receiving points. We assume that each primary receiving point can tolerate a maximum interference level. Also, secondary links have desired QoS performance in terms of BER.

We assumed a model in which $S$ base-stations are sharing $S$ different frequency bands. Each band has a user capacity of $K$ with throughput of $R$ per user. Therefore, each band can support an aggregate traffic of $KR$ bits per second per Hertz. In theory, each base-station achieves this throughput via the licensed band provided that there are $K$ active users and enough packets from each user to fully exploit the capacity.

We should note at this point that the proposed sharing protocol applies to both downlink and uplink transmissions. The access strategy used by each base-station to serve the users could be any of the standard techniques.

Active users are defined to be users of the wireless network requesting access for their data flow. Assume each base station $i \; \varepsilon \; \{1, 2, \ldots, S\}$ has a random number of requests for establishing a session say $a_i$ each distributed according to a Poisson distribution with average rate of $\lambda$. Each $a_i$ is assumed to be independent from requests at other stations. We assume that all users and sessions have the same data rate requirements met by the rate *R*.

## 4. Performance Analysis

Assume that there are *M* primary receiving points and *N* secondary communication links in the considered geographical area. Let us denote the channel gain from the transmitting node of secondary link *i* to receiving node of secondary link *j* by $g^s_{(j,i)}$ while the channel gain from the transmitting node of secondary link *i* to primary receiving point *j* as $g^p_{(j,i)}$.

If $N_i$ denotes the total noise and interference at the receiving side of secondary link *i*, for wireless access system, the corresponding effective bit-energy-to-noise spectral density ratio can be written as

$$\mu_i = \frac{W}{R_i} \frac{g^s_{(j,i)} P_i}{\sum_{j=1, j\neq i}^{N} g^s_{(j,i)} P_j + N_i}$$

Where *W* is the spectrum bandwidth, $R_i$ is the transmission rate of secondary link *i*. Here, $W/R_i$ is the processing gain which is usually required to be larger than a particular value. The processing gain is simply equal to one for other multiple access technologies and $\mu_i$ denotes the SINR. Now, if a particular modulation scheme is employed, there will be an explicit relation between BER and SINR.

Thus, for a specific required BER level of secondary link *i*, $\mu_i$ is required to be larger than a corresponding value $\gamma_i$. Hence, the QoS requirement for secondary link *i* can be expressed as

$$\mu_i \geq \gamma_i, i=1,2,...,N$$

Now, let $T_j$ be the minimum interference level tolerable by primary receiving point j. The interference constraint for primary receiving point *j* can be written as

$$\sum_{i=1}^{N} g^p_{(j,i)} P_i \leq T_j, j=1,2,...,M$$

where total interference at the primary receiving point *j* should be smaller the tolerable limit.

We will assume that transmission rate of secondary link *i* can be adjusted in an allowable range with minimum and maximum values are $R_i^{min}$, $R_i^{max}$ respectively. Also, power of secondary link *i* is constrained to be smaller than the maximum limit $P_i^{max}$.

## 5. Proposed Algorithm

The proposed work, QoS of Spectrum sharing among multiple Service Providers is carried out in a long-term spectrum Assignment scheme. The function coordinates and negotiates the spectrum assignments between multiple Service Providers for large geographical areas.

The spectrum assignments are updated periodically and it is explained with the help of the proposed algorithm named SBAC (Selection of Best Available Channel).

Algorithm : SBAC

$BS \leftarrow mn$ (mobile node send request to Base Station)
$CR \leftarrow BS$
$nCR \leftarrow CR$
$current\_channel\_available\_list \leftarrow nCR$
$prob \leftarrow current\_channel\_available / total\_channel$
$\forall ch\_frq$
    if $frmax < ch\_frq$
        $frmax \leftarrow frq$
    end
    if $frmin > ch\_frq$
        $frmin \leftarrow ch\_frq$
    end
end
$inter \leftarrow |frmax - frmin|$
$cost \leftarrow t*60*c$
$ch\_u = (10*\beta_1 * prob) + \beta_2 * \log \frac{1}{inter} + \beta_3 * \frac{1}{cost}$
$cu\_list[cu\_count++] = |ch\_u|$
if $(cu\_count != 0)$
    $\forall cu\_list$
        if $maxi < cu\_list$
            $maxi = ch\_list$
        end
    end
$channel\_maxi \leftarrow maxi$
end

## 6. Simulation Results

The Cognitive Radio receives the RF environment at its receiver and involves itself in a decision-making process to accommodate a new user requesting the spectrum allocation. This requires a decision-making considering certain factors, such as the secondary user's requirements as parameters like, its Channel coding, data transmission rate, etc. The user needs the spectrum to carry out its communications and specifies its QoS requirements to the cognitive radio that also gets the information about the RF environment from a sensing module.

The utility function represents the radio's behavioral traits for the decision-making process to achieve the required optimization. There can be many possible traits that can be considered in this regard but we shall consider only some of the basic traits for the radio in this research. Some of the possible traits that can be considered are the occupied bandwidth, spectral efficiency, throughput and delay.

We shall just consider a few parameters only, in order to maintain the simplicity in the research. These are the frequency bands, power and BER.

### 6.1 Minimize Propagation Delay

Propagation delay is the amount of time taken for the signal to travel from the sender to the receiver over a medium. It can be computed as the ratio between the link length and the propagation speed over the specific medium.

$$\text{Propagation delay} = d / s$$

where d is the distance travelled and s is the propagation speed.

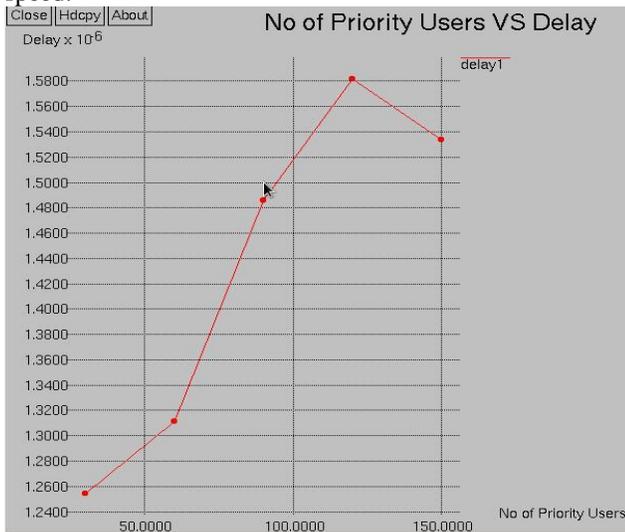

Fig : 1 No. of users VS Propagation Delay

In the proposed work, as the number of Users increases propagation delay gets minimized as shown in the Fig : 1.

### 6.2 Maximize Throughput

The maximum data rates of the TX and RX depend on the bandwidth of the circuits, and simplistically it might seem that the least of these will settle the issue.

But for communication between them the max data rate is affected by the noise in the system, and this will depend on the noise of the propagation medium, the noise figure of the RX, the power level of the TX, the transmission loss, and the maximum tolerable error rate.

Satisfactory data transmission can be achieved with higher noise at a lower bit rate because of the statistical nature of the noise, and the time-domain averaging of signals which occurs in the RX.

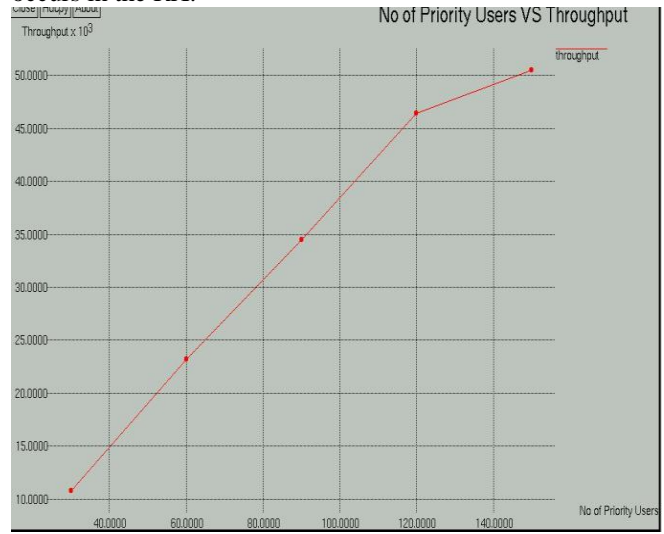

Fig : 2 No. of Users VS Throughput

The objective of the paper represents that improving the communications quality of the radio. Maximizing the throughput deals with the data throughput rate of the system. Emphasizing this objective, the overall system throughput should be increased and it is reached in the proposed work as shown in the Fig : 2. This refers to the increase in overall data throughput transmitted by the signal.

### 6.3 Minimize RTT

Round-trip time (RTT) is the time it takes for a client to send a request and the server to send a response over the network, not including the time required for data transfer.

Current Round-Trip Time (RTT) of every active connection is estimated in order to find a suitable value for the retransmission time-out.

RTT is the major contributing factor to latency on "fast" (broadband) connections and it's especially important to minimize the number of requests that the client needs to make and to parallelize them as much as possible.

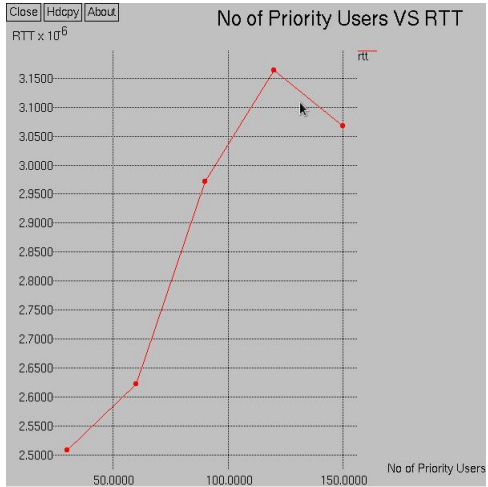

Fig : 3 No. of Users VS RTT

In the proposed work, as the number of priority users increases the RTT value decreases, as to minimize the number of round trips that need to be made.

### 6.4 Minimize Interference

Interference is the key factor that limits the performance of wireless networks.

As the mean call arrival increases the Interference gets decreased and there will be a minute variations in the Interference as shown in the Fig : 4.

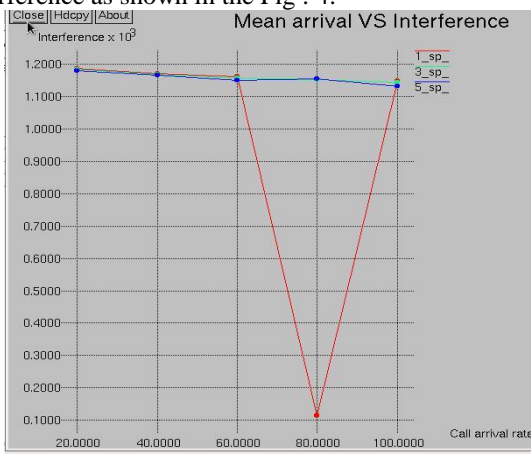

Fig : 4   Mean Call Arrival VS Interference

### 6.5 Maximize Spectral Efficiency($\eta$)

The Spectrum Efficiency $\eta_s$ is the ratio of average busy channels over total channels owned by service providers ie it refers to the amount of information that can be transmitted over a given bandwidth.

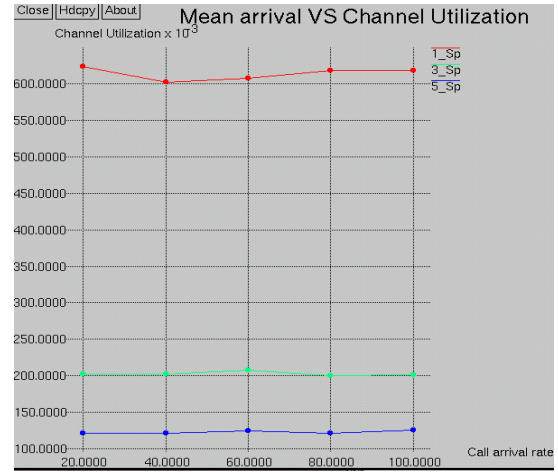

Fig : 5   Mean Call Arrival VS Spectrum Efficiency

As the mean call arrival increases the channel Utilization also increases as shown in the Fig :5. Higher Spectrum efficiency is estimated because the call blocking rate is lower; thus more calls can contribute to the spectrum utilization

## 7. Conclusions

The proposed dynamic spectrum sharing algorithm has been shown to be an effective solution for improving the spectrum efficiency under fluctuating traffic loads while maintaining the Interference and throughput in their acceptable QoS levels. It is illustrated that this model can be successfully employed in the key spectrum allocation decisions in such a spectrum sharing environment in a heterogeneous wireless network.

In this paper, we have studied the distribution of the interference generated by a secondary network to a primary network. We have derived a general formula for the interference taking into account the cognitive ability, throughput and transmit power. Also, Cognitive radio parameters, Utility Functions, QoS and interference constraint parameters on network performance are investigated and discussed.

**R.Kaniezhil** is a member of IEEE. She is a Research scholar in the Department of Computer Science, Periyar University, Salem. She received her B.Sc Degree from University of Madras in 1998. She received her MCA and M.Phil Degrees from Periyar University and Annamalai university, in 2001 and 2007, respectively. Her research interests include Mobile computing, Spectrum and Wireless Networking

**C.Chandrasekar** is a member of IEEE. He received his Ph.D degree from Periyar university. He is working as an Associate Professor, Department of Computer Science, Periyar University, Salem. His areas of interest include Wireless networking, Mobile Computing, Computer Communications and Networks. He is a research guide at various universities in India. He has published more than 40 technical papers at various National & International conferences and 43 journals.

**S.Nithya Rekha** is a member of IEEE. She is a Research scholar in the Department of Computer Science, Periyar University, Salem. She received her B.Sc Degree from Bharathiayar University in 1994. She received her MCA and M.Phil Degrees from IGNOU and PRIST university, in 2006 and 2008, respectively. Her research interests include Mobile Computing, Rough set and Wireless networking.